\documentclass{article}
\usepackage[T1]{fontenc} 
\usepackage[utf8]{inputenc} 
\usepackage{ismir,amsmath,cite,url}
\usepackage{graphicx}
\usepackage{tabularx}
\usepackage{booktabs}
\usepackage{multirow}
\usepackage{color}
\usepackage{amssymb}

\title{MusiConGen: Rhythm and Chord Control for Transformer-Based Text-to-Music Generation}

\multauthor
{Yun-Han Lan$^{1,2}$ \hspace{1cm} Wen-Yi Hsiao$^1$ \hspace{1cm} Hao-Chung Cheng$^2$ \hspace{1cm} Yi-Hsuan Yang$^{1,2}$} {$^1$ Taiwan AI Labs \hspace{1cm} $^2$ National Taiwan University\\
{\tt\small cyan0731@gmail.com, wayne391@ailabs.tw, \{haochung,yhyangtw\}@ntu.edu.tw}
}

\def\authorname{Y. Lan, W. Hsiao, H. Cheng and Y. Yang}

\begin{document}

\maketitle
\begin{abstract}
Existing text-to-music models can produce high-quality audio with great diversity. However, textual prompts alone cannot precisely control temporal musical features such as chords and rhythm of the generated music. To address this challenge, we introduce MusiConGen, a temporally-conditioned Transformer-based text-to-music  model that builds upon the pretrained MusicGen framework. Our innovation lies in an efficient finetuning mechanism, tailored for consumer-grade GPUs, that integrates automatically-extracted rhythm and chords as the condition signal. During inference, the condition can either be musical features extracted from a reference audio signal, or be user-defined symbolic chord sequence, BPM, and textual prompts. Our performance evaluation on two datasets---one derived from extracted features and the other from user-created inputs---demonstrates that MusiConGen can generate realistic backing track music that aligns well with the specified conditions. We open-source the code and model checkpoints, and provide audio examples online, \url{https://musicongen.github.io/musicongen_demo/}.
\end{abstract}

\section{Introduction}\label{sec:introduction}
The realm of text-to-music generation has seen significant progress over the recent years
\cite{musiclm, copet2023simple,huang2023noise2music,liu2023audioldm2,li2024jen,lin2023contentbased,Wu2023MusicControlNet,musicldm2024,mustango,musicmagus,tsai24ismir}. These models span various genres and styles, largely leveraging textual prompts to guide the creative process. 
There have been two primary methodological frameworks so far. The first employs \emph{Transformer} architectures to model audio tokens \cite{dhariwal2020jukebox} derived from pre-trained audio codec models \cite{soundstream, defossez2022highfi, kumar2023highfidelity}; noted examples include MusicLM \cite{musiclm} and MusicGen \cite{copet2023simple}. The second  employs \emph{diffusion} models to represent audio through spectrograms or audio features, such as AudioLDM 2 \cite{liu2023audioldm2} and JEN-1 \cite{li2024jen}.

Text-to-music generation model generally relies on the global textual conditions to guide the music generation process. 
Textual prompts serving as high-level conceptual guides, however, introduce a degree of ambiguity and verboseness into the music generation for describing the musical features \cite{Wu2023MusicControlNet}. This inherent vagueness poses a challenge in precisely controlling temporal musical features such as melody, chords and rhythm, which are crucial for music creation.
Building on the success of MusicGen-melody \cite{copet2023simple} in melody control, our focus now shifts to enhancing chord and rhythm control, aiming to create a more integrated approach to music generation that captures the full spectrum of musical elements.

\begin{table}
\resizebox{\columnwidth}{!}{
 \begin{tabular}{l|cc|c}
  \toprule
  \multirow{2}{*}{Model} & Chord & Rhythm & Do not need \\
  & control & control & reference audio\\
  \midrule
  Coco-Mulla\cite{lin2023contentbased} & $\surd$ & $\surd$ & \\
  Music ControlNet\cite{Wu2023MusicControlNet} & & $\surd$ & $\surd$\\
  \midrule
  Ours & $\surd$ & $\surd$ & $\surd$\\
  \bottomrule
 \end{tabular}
 }
 \caption{The comparison for conditions and condition type of related temporally-conditioned text-to-music models.}
 \label{tab:table1}
\end{table}

Table \ref{tab:table1}  tabulates two existing studies that have explored the incorporation of time-varying chord- and rhythm-related attributes in text-to-music generation.
Coco-Mulla \cite{lin2023contentbased} is a Transformer-based model that employs a large-scale, 3.3B-parameter MusicGen model, finetuned with an adapted LLaMA-adapter \cite{zhang2023llamaadapter} for chord and rhythm control. 
For rhythm control in particular, Coco-Mulla uses drum audio codec tokens extracted from a reference drum audio signal as a condition for guiding the music generation, thereby demanding \emph{reference audio} for control. 
While it is appropriate to assume the availability of such reference \emph{audio} in some scenarios, for broader use cases we desire to have a model that can take user-provided \emph{text}-like inputs as well, such as the intended beats-per-minute (BPM) value (for rhythm) and the chord progression as a series of chord symbols (for chords). This function is  not supported by Coco-Mulla.

The other model, Music ControlNet \cite{Wu2023MusicControlNet}, leverages a diffusion model architecture and the adapter-based conditioning mechanism of ControlNet \cite{zhang2023adding}  to manipulate text-like, symbolic melody, dynamics, and rhythm conditions.
This diffusion model creates a spectrogram based on the provided conditions, which is then transformed into audio using their pretrained vocoder. For musical conditions, a 12-pitch-class chromagram representation is used for the melody, combined with beat and downbeat probability
curves concatenation for rhythm control, and an energy curve to adjust the dynamic volume. However, Music ControlNet does not deal with chord conditions.

In view of the limits of the prior works, we introduce in this paper MusiConGen, a Transformer-based text-to-music model that applies temporal conditioning to enhance control over rhythm and chord.
MusiConGen is finetuned from the pretrained MusicGen framework \cite{copet2023simple}.
We design our temporal condition controls in a way that it supports not only musical features extracted from reference \emph{audio} signals, but also the aforementioned user-provided \emph{text-like} symbolic inputs such as BPM value and chord progression.
For effective conditioning of such time-varying features, we propose ``adaptive in-attention'' conditioning by extending the in-attention mechanism proposed in the MuseMorphose model \cite{wu2021musemorphose}.
Table \ref{tab:table1} includes a conceptual comparison of MusiConGen with existing models in terms of the conditions and their types.

In our implementation, we train MusiConGen 
on a dataset of \emph{backing track music} comprising 5,000 text-audio pairs obtained from YouTube. This training utilizes beat tracking and chord recognition models to extract necessary condition signals without the need for manual labeling. 
We note that rhythm and chord controls are inherently critical for backing tracks, for backing tracks often do not include the primary melody and their purpose is mainly to provide accompaniment for a lead performer. 

Moreover, instead of using the 
adapter-based finetuning methods 
\cite{zhang2023llamaadapter,zhang2023adding,peft},
we apply the  straightforward ``direct finetuning'' approach  to accommodate the domain shift from general instrumental music (on which MusicGen was trained) to the intended backing track music.
We leave the use of adapter-based finetuning as future work.
To make our approach suited for operations on consumer-grade GPUs, we propose a mechanism referred to as ``jump finetuning'' instead of finetuning the full MusicGen model.

We present a comprehensive performance study involving objective and subjective evaluation using two public-domain datasets, MUSDB18\cite{musdb18} and RWC-pop-100\cite{rwc-pop}.
Our evaluation demonstrates MusiConGen's enhanced ability to offer nuanced temporal control, surpassing the original MusicGen model in producing music that aligns more faithfully with the given conditions.

The contributions of this work are two-fold. First, to our best knowledge, this work presents the first Transformer-based text-to-music generation model that follows user-provided rhythm and chords conditions, requiring no reference audio signals.
Second, we present efficient training configuration allowing such a model to be built by finetuning the publicly-available MusicGen model with customer-level GPU, specifically 4x RTX-3090 in all  our experiments.
We open-source the code, checkpoint, and information about the training data of MusiConGen on GitHub.\footnote{\url{https://github.com/Cyan0731/MusiConGen}}

\section{Background}
\subsection{Codec Models for Audio Representation}\label{subsec:codec}
In contemporary music generation tasks, audio signals are typically compressed into more compact representations using two main methods: Mel spectrograms and codec tokens. Mel spectrograms provide a two-dimensional time-frequency representation, adjusting the frequency axis to the Mel scale to better align with human auditory perception. Codec tokens, on the other hand, are often residual vector quantization (RVQ) tokens that are encoded from audio signals by a codec model \cite{soundstream, defossez2022highfi, kumar2023highfidelity}. Following MusicGen, we employ in our work the Encodec\,(32k) \cite{defossez2022highfi} as the pretrained codec model to encode audio data at a sample rate of 32,000 Hz. This Encodec model comprises 4 codebooks, each containing 2,048 codes, and operates at a code frame rate $f_s$ of 50 Hz.

\subsection{Classifier-Free Guidance}\label{subsec:cfg}
Classifier-free guidance \cite{ho2021classifierfree} is a technique initially developed for diffusion models in generative modeling to enhance the quality and relevance of the outputs without the need for an external classifier. This approach involves training the generative model in both a conditional and an unconditional manner, combining the output score estimates from both methods during the inference stage. The mathematical expression is as 
$\nabla_{x} \log \tilde{p_{\theta}}(x\vert c) = (1-\gamma)\nabla_{x}\log p_{\theta}(x) + \gamma \nabla_{x}\log p_{\theta}(x\vert c)$.
Here, $\gamma$ represents the guidance scale, which adjusts the influence of the conditioning information. 
We perform a weighted average of $f_{\theta}(x, c)$ and $f_{\theta}(x)$ when sampling from the output logits.

\subsection{Pretrained MusicGen Model}\label{subsec:musicgen}
The pretrained model used in our study is a  MusicGen model with 1.5B parameters, equipped with melody control (i.e., MusicGen-Melody). The melody condition employs a chromagram of 12 pitch classes at a frame rate  $f_{\mathcal{M}}$, denoted as $\mathcal{M} \in \mathbb{R}^{T_{f_{\mathcal{M}}} \times 12 \times 1}$, derived from the linear spectrogram of the provided reference audio. 
For text encoding, the model leverages the FLAN-T5\cite{chung2022scaling} as a text encoder to generate conditioning text embeddings, represented as $\mathcal{T} \in \mathbb{R}^{T_{t5} \times d_{t5} \times 1}$. Both the melody and text conditions undergo linear projection into a $D$-dimensional space before being prepended to the input audio embedding. Regarding the input audio for training, audio signals are initially encoded into RVQ tokens, $X_{rvq} \in \mathbb{R}^{T_{f_s} \times 1 \times 4}$, using the pretrained Encodec model. These tokens are then formatted into a ``delay pattern'' \cite{copet2023simple}, maintaining the same sequence length. Subsequently, an embedding lookup table, $W_{emb} \in \mathbb{R}^{N \times D \times 4}$, where $N$ represents for numbers of codes in a codebook, is used to represent the associated codes, summing contributions from each codebook of $X_{rvq}$ to form the audio embedding $X_{emb} \in \mathbb{R}^{T_{f_s} \times D \times 1}$. The input representation is then fed to the self-attention layers via additive sinusoidal encoding.

\begin{figure*}
 \centerline{
 \includegraphics[width=2\columnwidth]{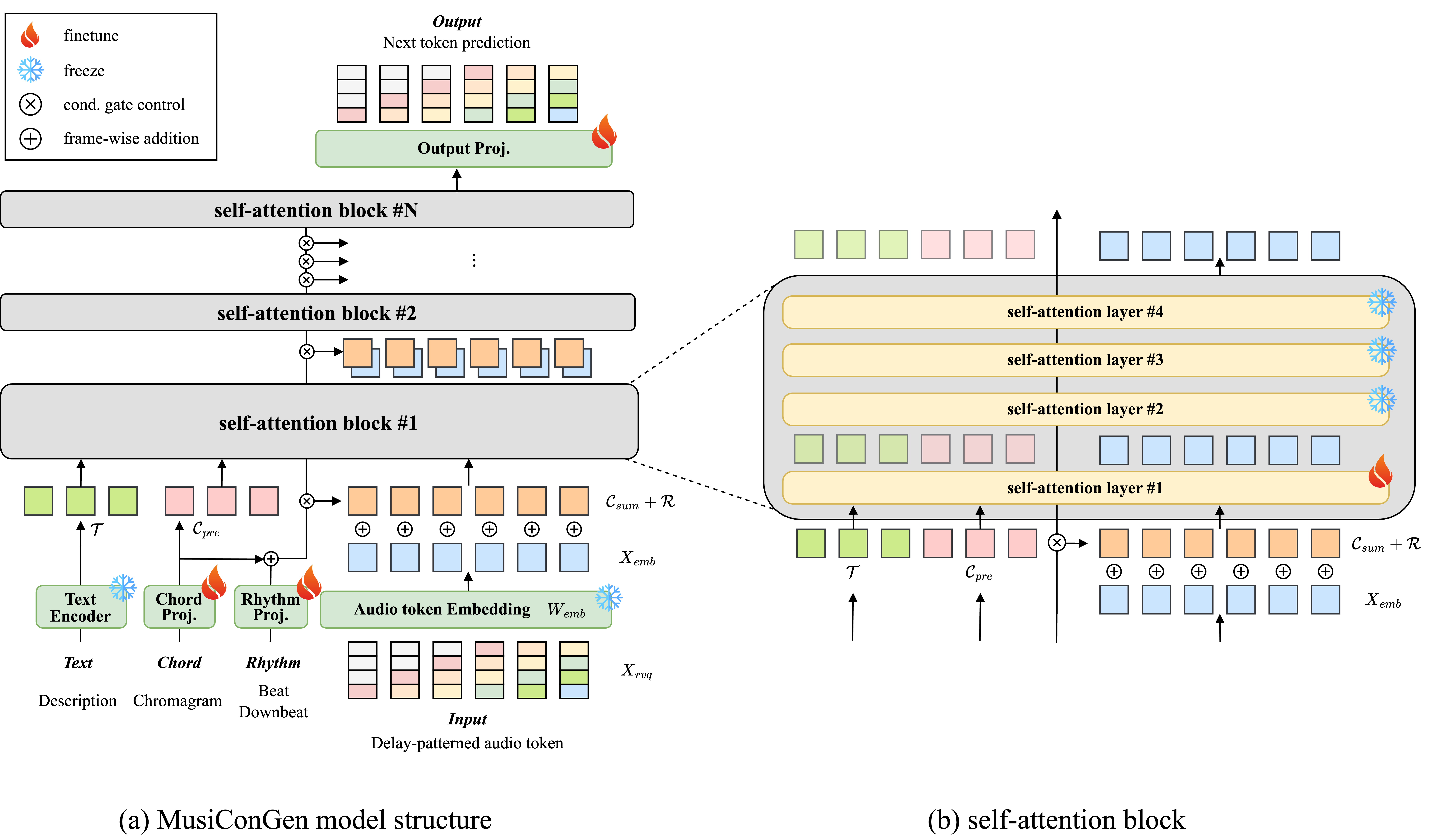}}
 \caption{The model structure of MusiConGen and the self-attention block. a) MusiConGen takes text $\mathcal{T}$, downsampled chord $\mathcal{C}_{pre}$ as prepended condition and frame-wise chord $\mathcal{C}_{sum}$ and rhythm $\mathcal{R}$ as additive condition. The addition operation of frame-wise conditions to each self-attention block is regulated by the condition gate control ($\otimes$). b) Each self-attention block consists of four layers. In our proposed model, only the first layer is finetuned, which is also called jump finetuning.}
 \label{fig:arch}
\end{figure*}

\section{Methodology}
Our method seeks to efficiently finetune the foundational MusicGen model using time-varying symbolic rhythm and chord conditions as guiding conditions. To achieve this, we must carefully consider both the representation of these conditions and the finetuning mechanism as follows:

\subsection{Representing Temporal \& Symbolic Conditions}\label{subsec:method_cond}

\textbf{Chords.} For chord condition, we employ two methods. The first \textbf{prepend} method is similar to the melody control method of MusicGen, denoted as $\mathcal{C}_{pre} \in \mathbb{R}^{T_{f_{\mathcal{M}}} \times 12}$ where $\mathcal{C}_{pre}$ maintains the same resolution (i.e. frame rate $f_{\mathcal{M}}$ and sequence length) as MusicGen's melody condition $\mathcal{M}$. This  allows us to utilize the pretrained melody projection weights from MusicGen as initial weights.
Furthermore, we have noted that chord transitions can lead to asynchronization issues. To address this, we introduce a second \textbf{frame-wise} chord condition, $\mathcal{C}_{sum} \in \mathbb{R}^{T_{f_s} \times 12 \times 1}$, which matches the resolution of the audio codec tokens, thus providing a solution for the synchronization problem.

\textbf{Rhythm.} To control rhythm, we derive conditions from both the beat and the downbeat. The beat represents the consistent pulse within a piece of music, and the downbeat signifies the first and most emphasized beat of each measure, forming the piece's rhythmic backbone. We encode beat and downbeat information into one-hot embedding each at a frame rate of $f_s$. For the beat embedding, a soft kernel is applied to allow for a tolerance of 70ms. Subsequently, the beat and downbeat arrays are summed to yield the \textbf{frame-wise} rhythm condition $\mathcal{R} \in \mathbb{R}^{T_{f_s} \times 1}$.

\subsection{Finetuning Mechanisms}\label{subsec:finetune}
The finetuning mechanism we employ consists of two parts: 1) jump finetuning, and 2) an adaptive in-attention mechanism. As illustrated in Figure \ref{fig:arch}, our proposed model activates condition gates at the ``block'' level, treating four consecutive self-attention layers as a block.

Jump finetuning is designed to specifically target the \emph{first} self-attention layer \emph{within each block} for finetuning, while freezing the remaining three self-attention layers of the same block, as shown in Figure \ref{fig:arch} (b). Doing so reduces the number of parameters of finetuning while maintaining the flexibility to learn to respond to the new conditions by refining the first self-attention layer per block.

The adaptive in-attention mechanism is designed to improve control over chords and rhythm. It is an adaptation of the in-attention technique of MuseMorphose \cite{wu2021musemorphose}, whose main idea is to augment every intermediate output of the self-attention layers with copies of the condition. Unlike the original implementation that augment all the self-attention layers, we selectively apply it to the first three-quarters of self-attention blocks (e.g., for a model with 12 blocks, in-attention is applied to first 9 blocks) to relax the control in the last few blocks for better balancing on rhythm and chords. This leads to better result empirically, as will be shown in Section \ref{subsec:eval2} and Table \ref{tab:table3}.

\section{Experimental Setup}
\subsection{Datasets}\label{subsec:dataset}
We finetuned the model using a dataset of $\sim$250 hours backing track music sourced from YouTube, comprising 5K songs across five genres: Rock, Funk, Jazz, Blues, and Metal, with 1K songs per genre. After preprocessing (see Section \ref{subsec:preproc}), the training data contained 80,871 clips. 

For evaluation, we used the rhythm and chords from two public-domain datasets---MUSDB18\cite{musdb18} and RWC-pop-100\cite{rwc-pop}. For MUSDB18, the rhythm and chords are \emph{extracted} from the audio signals, so this dataset reflects the case where the condition signals are from a \emph{reference audio}.  
There are 150 songs with four isolated stems: vocal, bass, drum, and others.
For each song, we dropped the vocals and divided the mix of the remaining tracks into 30-second clips, resulting in a total of 1,089 clips. 

The RWC comprises 100 Japanese pop songs with human annotated chord progressions and BPM labels. We simply use the \emph{human labels} as the conditions here, reflecting the case where the condition signals are user provided \emph{in a text-like format}.
We similarly divided each song into 30-second clips, leading to 755 clips in total. 

\subsection{Dataset Pre-processing Details}\label{subsec:preproc}
The training and evaluation datasets consist of full-song data, with durations ranging from 2 to 5 minutes per song. Below are the preprocessing details for each type of input:

\textbf{Audios:} All audio data have vocals removed. For the training and RWC dataset, we employed the source separation model Demucs\cite{rouard2022hybrid, defossez2021hybrid} to eliminate the vocal stem. In the MUSDB18 dataset, which already features isolated stems, we combined the bass, drum, and others stems to form the dataset. Each song was segmented into 30-second clips, ensuring each clip starts at a downbeat.

\textbf{Descriptions:} For the training set, the text prompts were simply extracted from the titles of the corresponding YouTube videos. For the two evaluation datasets, we tasked ChatGPT\cite{chatgpt2022} to generate 16 distinct text prompts, covering the five genres included by the training set. 
Here is an example---``A smooth acid Jazz track with a laid-back groove, silky electric piano, and a cool bass, providing a modern take on Jazz. Instruments: electric piano, bass, drums.'' At inference time, we randomly selected one of the 16 text prompts in a uniform distribution.

\textbf{Chords:} 
The RWC dataset comes with ground truth labeled chords.
For both the training set and MUSDB18, we used the BTC model \cite{Park2019ABT} as the chord extraction model to predict symbolic chords with time tags for each clip. The detailed chord quality extends to the seventh note. We then translated the extracted chord symbols with time tags into a 12-pitch chromagram in the order of C, C\#, ..., B. The chromagram's frame rate for the frame-wise condition $\mathcal{C}_{sum}$ is $f_s$, and for the prepend condition $\mathcal{C}_{pre}$ it is $f_{\mathcal{M}}$.

\textbf{Rhythm:} Except for RWC, beat and downbeat were extracted using the RNN+HMM model \cite{Bock2016d} from the Madmom library\cite{madmom}. The timing format for beats and downbeats was transformed into a one-hot representation matching the audio token frame rate $f_s$. A soft kernel was applied to the one-hot beat array to create a softened beat array. The rhythm representation $\mathcal{R}$ was the frame-wise summation of the softened beat array and downbeat array.

\begin{table*}[t]
 \begin{center}
 \begin{tabular}{@{}lc|c|ccc|cc@{}}
  \toprule
  \multirow{2}{*}{Model} & Evaluation & Rhythm & \multicolumn{3}{c|}{Chord}& \multirow{2}{*}{FAD} & \multirow{2}{*}{CLAP} \\ 
   & dataset & F-measure$(\%)$ & majmin$(\%)$ & triads$(\%)$ & tetrads$(\%)$ & & \\
  \midrule
  \textbf{proposed} &MUSDB18&\textbf{69.76}&67.03&66.19&56.91&\textbf{1.29}&0.34 \\
  ($\mathcal{C}_{pre}$$+$$\mathcal{C}_{sum}$$+$$\mathcal{R}$) &RWC    &\textbf{79.40}&\textbf{73.03}&\textbf{68.42}&\textbf{54.12}&\textbf{0.96}&0.34 \\
  \midrule
  \textbf{chords only}&MUSDB18&39.47&\textbf{73.25}&\textbf{72.29}&\textbf{60.89}&1.91&0.34 \\
  ($\mathcal{C}_{pre}$$+$$\mathcal{C}_{sum}$)                           &RWC    &49.85&\textbf{73.30}&\textbf{68.50}&50.66&2.18&0.34 \\
  \midrule
  \textbf{rhythm only}&MUSDB18&61.37&5.84&5.76&3.84&1.95&0.32 \\
  ($\mathcal{R}$) &RWC   &58.39&5.40&5.08&2.90&2.67&0.32 \\

  \midrule
  \textbf{no frame-wise chords} &MUSDB18&61.68&57.39&56.65&47.17&1.44&0.35 \\
  ($\mathcal{C}_{pre}$$+$$\mathcal{R}$) &RWC&69.30&60.95&57.19&44.21&1.29&0.35 \\
  \midrule
  \textbf{baseline} &MUSDB18&26.14&53.13&52.31&44.83&2.01&0.34 \\
  (no finetuning; $\mathcal{M}$ for $\mathcal{C}_{pre}$) &RWC    &30.67&51.90&48.54&35.81&2.30&0.35 \\
  
  \bottomrule
 \end{tabular}
 \end{center}
 \caption{Objective evaluation results for models with different conditions on two different test sets MUSDB18 and RWC. 
 With the proposed condition representation, we can achieve better performance both in rhythm and chord controls. }
 \label{tab:table2}
\end{table*}

\begin{table*}[t]
 \begin{center}
 \begin{tabular}{@{}lc|c|ccc|cc@{}}
  \toprule
  \multirow{2}{*}{Model} & Evaluation & Rhythm & \multicolumn{3}{c|}{Chord}& \multirow{2}{*}{FAD} & \multirow{2}{*}{CLAP} \\ 
   & dataset & F-measure$(\%)$ & majmin$(\%)$ & triads$(\%)$ & tetrads$(\%)$ & & \\
  \midrule
  \textbf{proposed}&MUSDB18&\textbf{69.76}&67.03&66.19&56.91&\textbf{1.29}&0.34 \\
   (jump$+$adaptive in-attn)                        &RWC    &\textbf{79.40}&\textbf{73.03}&\textbf{68.42}&\textbf{54.12}&\textbf{0.96}&0.34 \\
  \midrule

  \textbf{ablation 1}&MUSDB18&42.28&\textbf{71.06}&\textbf{70.21}&\textbf{61.58}&1.39&0.36 \\
  (jump finetuning only) &RWC    &53.14&\textbf{76.04}&\textbf{71.33}&\textbf{57.52}&1.27&0.36 \\
  \midrule
  \textbf{ablation 2}
        &MUSDB18&67.23&66.47&65.60&56.37&1.59&0.35 \\
  (jump$+$full in-attn)       &RWC    &71.13&64.82&60.77&48.07&1.47&0.35 \\
  \midrule
  \textbf{finetuned baseline}&MUSDB18&40.15&55.65&54.88&45.52&1.94&0.36 \\
  (jump only; no $\mathcal{C}_{sum}$ no $\mathcal{R}$)                         &RWC    &49.25&56.49&52.66&38.07&2.24&0.36 \\
  \bottomrule
 \end{tabular}
 \end{center}
 
 \caption{Objective evaluation results for models trained with different finetuning mechanisms. We see that the proposed jump finetuning with  adaptive (partial) in-attention achieves better result on rhythm and chord controls.}
 \label{tab:table3}
\end{table*}

\subsection{Training Configuration}\label{subsec:confg}
The proposed rhythm and chord-conditioned Transformer was built upon the architecture of the medium-sized (1.5B) MusicGen-melody, featuring $L=48$ self-attention layers with dimension $D=1,536$ and 24 multi-head attention units. The condition dropout rate is 0.5 and guidance scale is set to be $\gamma = 3$ for classifier-free guidance.
We finetuned only a quarter of the full model, which corresponds to 352 million parameters, while keeping both the audio token embedding lookup table and the FLAN-T5 text encoder frozen. The training involved 100K finetuning steps, carried out over approximately 2 days on 4 RTX-3090 GPUs, with a batch size of 2 per GPU for each experiment.

\subsection{Objective Evaluation Metrics}\label{subsec:evalmetric}
\begin{figure*}
 \centerline{
 \includegraphics[width=\textwidth]{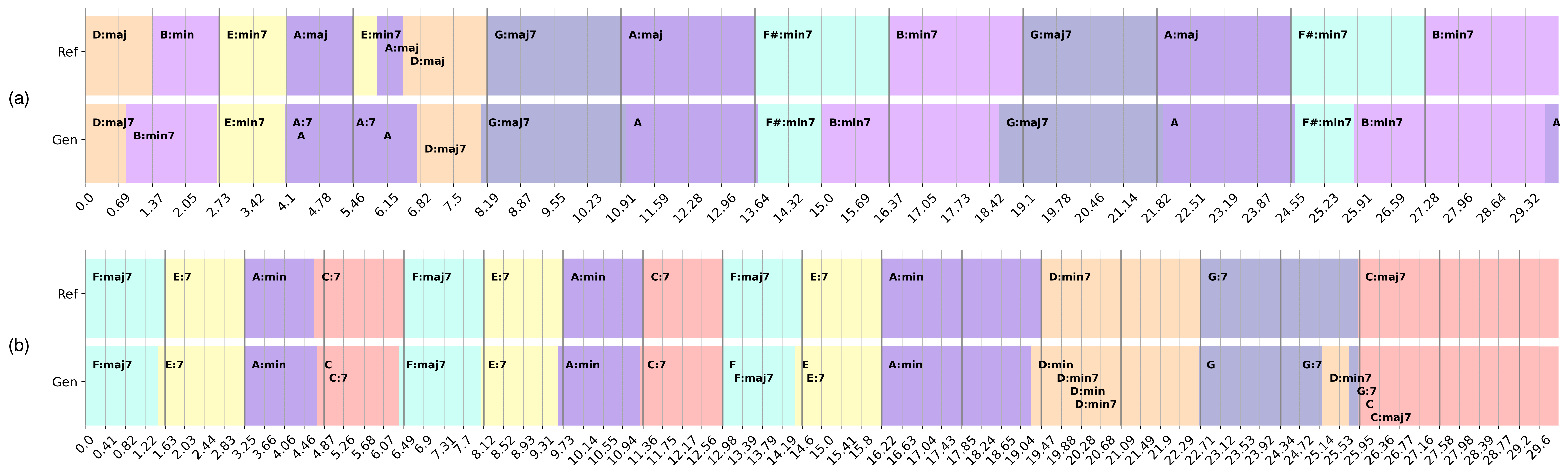}}
 \caption{Comparison on chord progression and beats of ground truth and generated samples, using the conditions from RWC. For each example (a) or (b), the top row is ground truth chords and the bottom row is extracted chords from generated samples. The thick and light gray lines indicate the times of the downbeat and the beat, respectively.}
 \label{fig:chord_comp}
\end{figure*}

We employed metrics to evaluate controllability of chords and rhythm, textual adherence and audio fidelity. For the first two metrics, we used the rhythm and chord conditions from a clip in a evaluation dataset to generate music (along with a text prompt generated by ChatGPT; see Section \ref{subsec:preproc}), applied the Madmom and BTC models on the generated audio to estimate beats and chords, and  evaluated how they reflect the given conditions.
See Figure \ref{fig:chord_comp} for  examples.

\textbf{Chord}. 
We used the \textit{mir\_eval} \cite{RaffelMHSNLE14} package to measure 3 different degrees of frame-wise chord correctness: \textbf{majmin}, \textbf{triads} and \textbf{tetrads}. The majmin function compares chords in major-minor rule ignoring chord qualities outside major/minor/no-chord. The triads function compares chords along triad (root \& qulaity to \#5), while the tetrads compares chords along tetrad (root \& full quality).

\textbf{Rhythm} F1 measurement follows the standard methodology for beat evaluation. We measured the beat accuracy also via \textit{mir\_eval}, assessing the alignment between the beat timestamps of the generated music and the reference rhythm music data, with a tolerance window of 70ms. 

\textbf{CLAP}\cite{laionclap2023, htsatke2022} score examines the textual adherence by  the cosine similarity between the embedding of the text prompt and that of the generated audio  in a text-audio joint embedding space learned by contrastive learning. 
Here, we used the LAION CLAP model trained for music \cite{laion-clap},
\texttt{music\_audioset\_epoch\_15\_esc\_90.14.pt}.

\textbf{FAD} is the Fr\'echet distance between the embeddings distribution from a set of reference audios and that from the generated audios \cite{kilgour2019frechet,gui2024adapting}. The metric represent how realistic the generated audios are compared to the given reference audios. The audio encoder of FAD we used is VGGish \cite{hershey2017cnn} model which trained on an audio classification task. The reference set of audios was from MUSDB18 or RWC depending on the evaluation set.

\subsection{Subjective Evaluation Metrics}\label{subsec:evalmetric2}

We also did a listening test to evaluate the followings aspects: text relevance, rhythm consistency, and chord relevance. Text relevance concerns how the generated audio clips reflect the given text prompts. Rhythm consistency is about how steady the beats is within an audio clip. (We found that, unlike the case of objective evaluations, minor out-of-sync beats at the beginning of a clip were deemed acceptable here perceptually.) Chord relevance concerns how a generated clip follows the given chord progressions. 

\section{Experimental Results}


\subsection{Objective Evaluation: Temporal Conditions}\label{subsec:eval1}
We assessed the audio generated under various condition combinations applied to the training model, including the proposed method and its \textbf{ablations} with either chord- or rhythm-only as the temporal condition, or using both but without the frame-wise chord condition. The finetuning configurations and mechanisms for these models were the same.
Moreover, we considered the \textbf{baseline} as follows.
The pretrained MusicGen-melody model originally processes text and melody conditions $\mathcal{T}, \mathcal{M}$. 
We simply used the prepend chord condition $\mathcal{C}_{pre}$ as input to the linear projection layers originally pretrained to take the melody condition, without finetuning the entire model at all.
In addition, we appended to the end of the text prompt BPM information (e.g., ``at BPM 90'') as the rhythm condition.

Result shown in Table \ref{tab:table2} leads to many findings. Firstly, a comparison between the result of the proposed model (first row) and the baseline (last row) demonstrates nicely the effectiveness of the proposed design. The proposed model leads to much higher scores in almost all the metrics.  Moreover, it performs similarly well for the two evaluation datasets, suggesting that MusiConGen can deal with both conditions extracted from a reference audio signals or provided by creators in a symbolic text-like format.

Secondly, although the baseline model does not perform well, it still exhibits some level of chord control, showing the knowledge of melody can be transferred to chords.

Finally, from the ablations (middle three rows), chord-only and rhythm-only did not work well for rhythm and chord control respectively, which is expected. 
Compared to the proposed model, excluding per-frame chord condition degrades both chord and rhythm controllability, showing that chord and rhythm are interrelated.

\subsection{Objective Evaluation: Finetuning Mechanisms}\label{subsec:eval2}
Besides the proposed finetuning method, we evaluated the following alternatives. 
\textbf{Finetuned baseline} is a baseline model that was finetuned using the prepended chords ($\mathcal{C}_{pre}$) instead of melody $\mathcal{M}$  the frame-level conditions, employing the jump finetuning mechanism but no in-attention.
\textbf{Jump finetuning without in-attention} (abalation 1)
and \textbf{jump finetuning with full in-attention} (abalation 2) are ablations which use full conditions (prepended chord $\mathcal{C}_{pre}$, frame-wise chord $\mathcal{C}_{sum}$, and rhythm $\mathcal{R}$), but we either dropped in-attention entirely, or employed in-attention to every self-attention block, instead of only the first three-quarter blocks as done by the proposed method.

The result is tabulated in Table \ref{tab:table3}. 
Among the four methods, the proposed method leads to the best rhythm control and very competitive chord control. 
Comparing the results of the proposed method and the two ablations reveals a trade-off in rhythm and chord control when we go from no in-attention, adaptive (partial) in-attention, to full in-attention.
The proposed method strikes an effective balance between rhythm and chord controls.


Comparing the last row of Table \ref{tab:table2} and that of Table \ref{tab:table3} shows that the finetuned baseline outperforms the baseline (with no finetuning at all) mainly in the rhythm control.
This is notable as the finetuned baseline is actually trained with only the prepend chord condition $\mathcal{C}_{pre}$, not using the rhythm condition $\mathcal{R}$, suggesting again the interrelation of chord and rhythm.
Moreover, although the finetuned baseline is better than the baseline, it is still much inferior to the proposed method in both chord and rhythm controls.

\begin{figure}
 \centerline{
 \includegraphics[width=1.1\columnwidth, trim={1cm 1cm 1cm 1cm}]{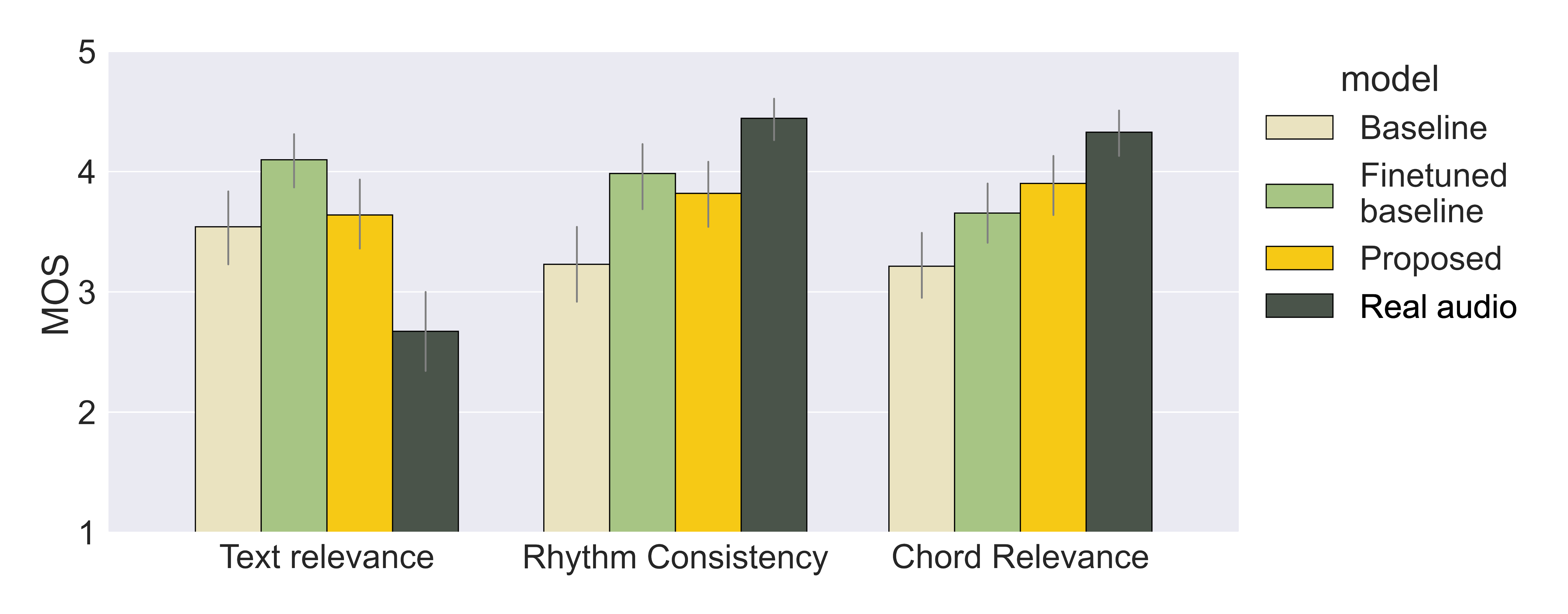}}
 \caption{Subjective evaluation of condition controls---\\5-scale mean opinion score  with 95\% confidence interval.}
 \label{fig:survey}
\end{figure}

\subsection{Subjective Evaluation}\label{subsec:survey}

We evaluated three models in the listening test: the baseline, the finetuned baseline, and the proposed model. 
Each model generates a music clip using the ChatGPT-generated text prompts, along with the BPM and chords from the RWC dataset, namely considering text-like symbolic rhythm and chord conditions. 
Besides the audios generated by the three models, we also included real audios from the RWC dataset as the \textbf{real audio}. We note that the real audios would have perfect rhythm and chord controllability (for they are where the conditions are from), but the textual adherence would be bad because RWC songs are J-Pop rather than any of the five genres (i.e., Rock, Funk, Jazz, Blues, and Metal) described by the text prompts.

We had 23 participants in the user study, 85\% of whom have over three years of musical training. 
Each time, we displayed the given text, rhythm and chord conditions, and asked a participant to rate the generated audio and the real audio (anonymized and in random order) on a five-point Likert scale. 
The  result is shown in Figure \ref{fig:survey}.

Several findings emerged. Firstly, the proposed model demonstrated superior chord control compared to the other two models, although it still fell short of matching the real audio. Secondly, the proposed model has no significant advantage on rhythm consistency against the finetuned baseline. As suggested by the examples on our demo page, we found that being on the precise beat onset does not significantly impact rhythm perception. Thirdly, our model had lower text relevance than the finetuned baseline, suggesting that our model may have traded text control for increased temporal control of rhythm and chords.

\section{Conclusion and Future Work}

This paper has presented conditioning mechanisms and finetuning techniques to adapt MusicGen for better rhythm and chord control. 
Our evaluation on backing track generation shows that the model can take condition signals from either a reference audio or a symbolic input.
For future work, our user study shows room to further improve the rhythm and chord controllability while keeping the text relevance. 
This might be done by scaling up the model size, better language model, or audio codecs.
It is also interesting to incorporate
additional conditions, such as symbolic melody, instrumentation, vocal audio, and video clips.

\section{Acknowledgements}
We are grateful to the discussions and feedbacks from the research team of Positive Grid, a leading global guitar amp and effect modeling company, during the initial phase of the project. 
We also thank the comments from the anonymous reviewers and meta-reviewer.
The work is also partially supported by grants from the National Science and
Technology Council (NSTC\,112-2222-E-002-005-MY2), (NSTC\,113-2628-E-002-029), and from the Ministry of Education of Taiwan (NTU-112V1904-5).

\bibliography{reference}

\end{document}